\documentclass{article}
\usepackage{graphicx}
\usepackage[margin=1.5cm]{geometry}
\usepackage[utf8]{inputenc}
\usepackage{longtable}
\usepackage{url}
\usepackage{algorithm}
\usepackage{algorithm} 
\usepackage{algorithmicx}
\usepackage{algpseudocode}
\usepackage{amsmath} 
\usepackage{array}
\usepackage{authblk}
\usepackage{hyperref}
\usepackage{cleveref}
\usepackage{adjustbox}
\usepackage{url}

\title{Autono: A ReAct-Based Highly Robust Autonomous Agent Framework\footnote{Repo: \url{https://github.com/vortezwohl/Autono}}}
\author{Zihao Wu \thanks{Email: vortez.wohl@gmail.com}}
\affil{Fujian University of Technology}
\date{\today}
\renewcommand{\algorithmicrequire}{ \textbf{Input:}}
\renewcommand{\algorithmicensure}{ \textbf{Output:}}

\begin{document}

\maketitle

\begin{abstract}
    \noindent
    This paper proposes a highly robust autonomous agent framework based on the ReAct paradigm, designed to solve complex tasks through adaptive decision making and multi-agent collaboration. Unlike traditional frameworks that rely on fixed workflows generated by LLM-based planners, this framework dynamically generates next actions during agent execution based on prior trajectories, thereby enhancing its robustness. To address potential termination issues caused by adaptive execution paths, I propose a timely abandonment strategy incorporating a probabilistic penalty mechanism. For multi-agent collaboration, I introduce a memory transfer mechanism that enables shared and dynamically updated memory among agents. The framework's innovative timely abandonment strategy dynamically adjusts the probability of task abandonment via probabilistic penalties, allowing developers to balance conservative and exploratory tendencies in agent execution strategies by tuning hyperparameters. This significantly improves adaptability and task execution efficiency in complex environments. Additionally, agents can be extended through external tool integration, supported by modular design and MCP protocol compatibility, which enables flexible action space expansion. Through explicit division of labor, the multi-agent collaboration mechanism enables agents to focus on specific task components, thereby significantly improving execution efficiency and quality. 
\end{abstract}

\section{Introduction}

In recent years, the rapid advancement of artificial intelligence technology has revolutionized the field of natural language processing (NLP), enabling the development of increasingly sophisticated autonomous agent frameworks. These frameworks aim to solve complex tasks by leveraging the capabilities of large language models (LLMs) and multi-agent collaboration mechanisms. Starting with the introduction of the Transformer model\cite{vaswani2017attention}, which laid the foundation for modern deep learning architectures, the field has witnessed groundbreaking innovations such as the pre-trained BERT model\cite{devlin2018bert}, chain-of-thought (CoT) prompting\cite{wei2022cot}, and the ReAct paradigm\cite{yao2022react}. Each of these advancements has contributed significantly to enhancing the reasoning, adaptability, and efficiency of AI systems in handling complex tasks.

Despite these achievements, existing autonomous agent frameworks still face several limitations in flexibility, robustness, and multi-agent collaboration\cite{talebirad2023multi}. Traditional frameworks, such as AutoGen and LangChain, often rely on predefined workflows generated by LLM-based planners\cite{huang2024llmplan}. While these approaches work well in structured environments, they struggle to adapt to uncertainties and dynamic changes in complex and real-world scenarios. This rigidity can lead to suboptimal performance, resource inefficiencies, and even task failures when faced with unexpected challenges. Furthermore, multi-agent collaboration mechanisms in existing frameworks often lack effective mechanisms for information sharing and dynamic coordination, limiting their ability to handle complex tasks that require seamless interaction among multiple agents.

To address these limitations, this paper proposes a highly robust autonomous agent framework based on the ReAct paradigm. The ReAct paradigm\cite{yao2022react}, which integrates reasoning and action in an iterative process, has demonstrated significant advantages in reducing hallucinations\cite{huang2023hallucination} and improving task accuracy by allowing agents to interact with their environment and refine their reasoning in real time. By dynamically generating next actions during agent execution based on prior trajectories, my framework enhances flexibility and adaptability, enabling agents to respond effectively to dynamic task requirements and environmental feedback.

A key innovation of this framework is the introduction of a timely abandonment strategy with a probabilistic penalty mechanism. This strategy dynamically adjusts the probability of task abandonment based on task progress and resource consumption, ensuring timely termination of unproductive or overly resource-intensive tasks. By tuning hyperparameters such as abandonment probability and penalty coefficient, developers can balance conservative and exploratory tendencies in agent execution strategies, thereby improving adaptability and task efficiency in complex environments.

Additionally, the framework implements a memory transfer mechanism that enables agents to share and dynamically update information during multi-agent collaboration. This mechanism allows agents to focus on specific task components while maintaining a shared understanding of the overall task context, significantly improving execution efficiency and quality. Furthermore, the framework adopts a modular design approach, enabling users to customize and optimize agent behaviors according to specific requirements. This design enhances the framework's applicability and practicality while reducing development and deployment complexity.

The proposed framework also supports the Model Context Protocol (MCP)\cite{hou2025mcp}, an open specification designed to standardize how context is passed between LLMs and applications. By providing a structured framework for communication, MCP enhances reliability, security, and functionality, enabling seamless integration with various tools and data sources. This compatibility further extends the framework's versatility and applicability in diverse scenarios.

In summary, this paper aims to address the limitations of existing agent frameworks by proposing a novel and highly robust autonomous agent framework. By leveraging the ReAct paradigm, timely abandonment strategy, memory transfer mechanism, and MCP compatibility, the framework is designed to enhance adaptability, robustness, and efficiency in solving complex tasks. The contributions of this paper are expected to advance the state of the art in autonomous agent frameworks.

\section{Related Work}

In the flourishing field of artificial intelligence, language models have achieved remarkable advances in natural language processing. Starting with the groundbreaking Transformer model, the field has seen a series of innovations, including the pre-trained BERT model, chain-of-thought (CoT) prompting, the ReAct paradigm, and multi-agent collaboration mechanisms.

Chain-of-Thought (CoT) prompting\cite{wei2022cot} is a technique that enables reasoning capabilities through intermediate steps. It helps large language models (LLMs) improve their performance in tasks requiring logical progression, such as mathematical and logical problems\cite{wei2022cot}. By explicitly describing intermediate reasoning steps, CoT enhances the transparency and precision of AI’s thought process and can be combined with few-shot prompting\cite{10147172} for better results on complex tasks with limited examples.

ReAct (Reasoning + Acting)\cite{yao2022react} is a prompt engineering method that integrates reasoning and action in an iterative process. It allows AI to interact with an environment, learn from new data, and refine its reasoning in real time. The method works by generating reasoning about a problem, interacting with an environment (e.g., querying APIs, searching the web), and using the results to inform the next reasoning step, creating a feedback loop. This combination of CoT with a connection to external information sources significantly reduces hallucinations, making ReAct agents more accurate and trustworthy.

Multi-Agent Collaboration Mechanisms\cite{talebirad2023multi} have also evolved to address complex tasks that require coordination among multiple agents. These mechanisms enable agents to work together, share information, and achieve common goals.

Notably, the AutoGen framework\cite{wu2023autogen}, designed as an advanced multi-agent dialogue system, supports the creation and collaboration of multiple agents. However, its task termination depends primarily  on preset maximum steps or human feedback, lacking a flexible, adaptive task termination mechanism.

Similarly, the CrewAI framework\cite{CrewAI_crewAI_2024} emphasizes role-playing and collaboration among agents but falls short in dynamic task execution, struggling to respond effectively to sudden changes during task execution. The LangChain framework\cite{chase2022langChain} offers fine-grained control but is also limited in dynamically adjusting execution paths based on real-time task status. The Swarm framework\cite{OpenAI_Swarm_2023}, primarily used for educational purposes, is lightweight but lacks production-grade features and flexibility, making it insufficient for real-world complex tasks. The Magnetic-One framework\cite{fourney2024magentic} explores a novel multi-agent collaboration architecture to overcome the limitations of single large models in handling complex tasks but faces challenges in practical applications, such as the need for efficient communication and coordination mechanisms among agents, and potential computational and communication bottlenecks in large-scale multi-agent systems.

Moreover, none of the above frameworks supports the Model Context Protocol (MCP). MCP is an open specification designed to standardize how context is passed between LLMs and applications\cite{hou2025mcp}. MCP provides a structured framework that improves reliability, security, and functionality by addressing challenges such as standardizing communication, enhancing context management, improving security, enabling complex workflows, and promoting interoperability. MCP defines a structured message format\cite{hou2025mcp} that separates different types of information being sent to and from language models, including system messages, user messages, assistant messages, tool messages, and metadata. This structured approach allows for clearer delineation of roles and responsibilities within the communication flow, reducing ambiguity and potential confusion for the model.

One of the key advantages of MCP is its plug-and-play characteristic, platform independence, and technology stack independence. MCP acts as a universal interface, enabling developers to avoid the fragmentation of maintaining separate integrations. Once an application supports MCP, it can connect to any number of services through a single mechanism. This dramatically reduces the manual setup required each time you want your AI to use a new API. Teams can focus on higher-level logic rather than reinventing connection code for the 10th time. MCP replaces fragmented integrations with a simpler, more reliable single protocol for data access\cite{hou2025mcp}, allowing AI applications to seamlessly connect to various tools and data sources without the need for custom coding each from scratch.

To address these limitations, this paper proposes a highly robust autonomous agent framework based on the ReAct paradigm. The framework introduces several innovative mechanisms to enhance adaptability and robustness in complex environments. One of the key innovations of this framework is its compatibility with the MCP protocol, which enables seamless integration with various tools and data sources, enhancing the framework's versatility and applicability in diverse scenarios.


\section{Key Concepts}

This framework is built around two key concepts: \textbf{Agent} and \textbf{Tool}.

\noindent An agent is defined as an autonomous artificial intelligence entity with the following properties: 

\begin{enumerate}
    \item \textbf{Autonomy:} Agents can execute tasks independently without human intervention, capable of self-action and self-correction.
    
    \item \textbf{Goal-Driven:} All actions of the agent are ultimately aimed at achieving specific goals.

    \item \textbf{Extensible:} The agents' capabilities can be directly extended without modifying its underlying structure.

    \item \textbf{Limited Action Space:} An agent only executes actions within its defined action space.

    \item \textbf{Infinite State Space:} The agents' states are derived from action feedback, which is determined by the environment.

    \item \textbf{Observable:} The agents' operations are observable and modifiable, allowing users to intervene in its action trajectory.

    \item \textbf{Memory Capability:} The agents can remember its trajectories, and different agent instances can share and transfer memories, enabling multi-agent collaboration.

    \item \textbf{Controllable Behavior:} An agent's behavior tendencies can be controlled, which allowing users to adjust the agent's inclination toward cautiousness or exploration.
\end{enumerate}

\noindent A tool is an encapsulated action logic that the agent can invoke, with the following properties:

\begin{enumerate}
    \item \textbf{Descriptive:} Tools have descriptive metadata, such as name, description, parameter requirements, and return description.

    \item \textbf{Parameterized:} Tools can receive input parameters that define their execution manner and conditions.

    \item \textbf{Semantic Feedback:} The response of a tool is semantically rich, typically in the form of text or a string, indicating the execution result and environmental feedback.
\end{enumerate}

\noindent
Tools are divided into general tools and handoff tools. General tools are common tools an agent can use, while handoff tools enable task handover and memory transfer between agents. When an agent encounters a subtask it cannot solve, it checks its handoff tools and passes the task to another agent with the appropriate capabilities.

\newpage

\section{System Design}

\begin{figure}[htbp]
    \centering
    \includegraphics[width=0.8\textwidth,keepaspectratio]{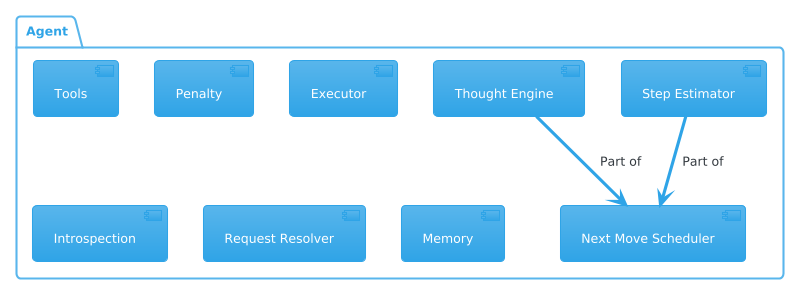}
    \caption{Components of an Agent}
    \label{fig:agent_components}
\end{figure}

\Cref{fig:agent_components} shows the architecture design of the agent. An agent consists of the following components:

\begin{enumerate}
    \item \textbf{Thought Engine:} The brain of the agent, typically a fine-tuned general language model.
    
    \item \textbf{Tools:} Encapsulated action logic that defines the agent's action space.
    
    \item \textbf{Step Estimator:} Estimates the number of steps required to complete a task based on the agent's action space.
    
    \item \textbf{Penalty:} Implements the probabilistic penalty mechanism for the timely abandonment strategy.
    
    \item \textbf{Memory:} Stores the agent's execution trajectories and state changes, enabling short-term memory and memory transfer between agents.
    
    \item \textbf{Request Resolver:} Understands tasks, decomposes subtasks, and extracts goals.
    
    \item \textbf{Next Move Scheduler:} Estimates the next action based on the agent's action trajectories and state changes, selecting the appropriate ability and generating corresponding parameters.
    
    \item \textbf{Executor:} Executes the actions, analyzes the results and environmental feedback, and updates the agent's trajectory.
    
    \item \textbf{Introspection:} Summarizes the agent's actions and extracts task results after the agent reaches its final state or exits due to the timely abandonment strategy.
\end{enumerate}

\begin{figure}[htbp]
    \centering
    \includegraphics[width=0.3\textwidth,keepaspectratio]{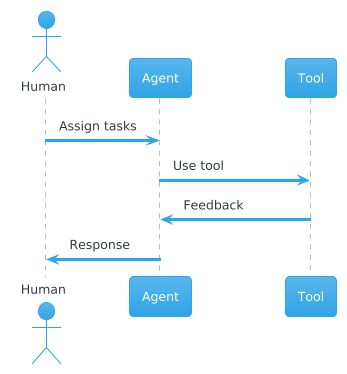}
    \caption{Task Execution Workflow of an Agent}
    \label{fig:task_execution}
\end{figure}

\newpage
\noindent
\Cref{fig:task_execution} illustrates the process of a task execution workflow. It involves three entities: Human, Agent, and Tool. The process begins with the Human assigning tasks to the Agent. Upon receiving the task, the Agent uses a Tool to perform the task. After the Tool completes the task, it sends feedback to the Agent. Finally, the Agent responds to the Human with the results of the task execution. This sequence of actions demonstrates how an agent interacts with humans and tools to accomplish tasks. 

\begin{figure}[htbp]
    \centering
    \includegraphics[width=1.0\textwidth,keepaspectratio]{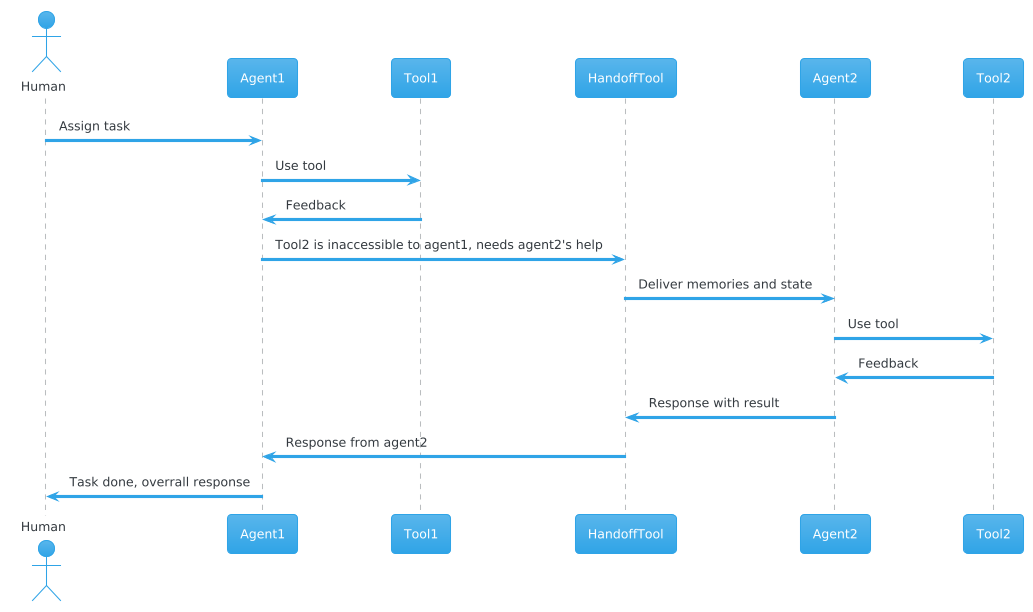}
    \caption{Task Execution Workflow Involving Multiple Agents}
    \label{fig:multi_agent_task_execution}
\end{figure}

\noindent
\Cref{fig:multi_agent_task_execution} illustrates the process of a task execution workflow involving multiple agents. It involves four entities: Human, Agent1, Tool1, HandoffTool, Agent2, and Tool2. The process begins with the Human assigning a task to Agent1. Upon receiving the task, Agent1 uses Tool1 to perform the task. After Tool1 completes its part, it sends feedback to Agent1. However, when Agent1 encounters a subtask that it cannot solve (e.g., Tool2 is inaccessible to Agent1), it uses the HandoffTool to pass the task and its memories to Agent2. Agent2 receives the task and memories, uses Tool2 to complete the task, and sends feedback to Agent1. Finally, Agent1 responds to the Human with the overall results of the task execution. This sequence demonstrates how multiple agents interact with humans, tools, and handoff mechanisms to collaboratively accomplish complex tasks.


\section{MCP Compatibility}

To ensure compatibility with the MCP protocol, I design and implement an MCP tool adapter and session management mechanism. The MCP protocol, a standardized communication protocol, aims to address the interaction challenges between large language models and external tools, providing flexible tool invocation capabilities for agents.

\paragraph{Tool Adapter}

The core idea of the MCP tool adapter is to encapsulate MCP-defined tool interfaces into tools that agents can directly invoke. The adapter parses the tool description information defined by the MCP protocol to dynamically generate metadata for tools, including tool names, parameter types, and descriptions. This encapsulation allows agents to invoke tools from different sources in a unified manner without concerning themselves with the specific implementation details of the tools.

\paragraph{Clients}

The MCP client is responsible for establishing connections with MCP servers and managing the communication process. I implement multiple client adapters, including those based on standard input/output, Server-Sent Events (SSE), and WebSocket, to support different types of MCP servers. The client design adopts a modular approach, dynamically selecting the appropriate communication method based on the specific MCP server type.

\paragraph{Session Management}

To ensure efficient and reliable communication between agents and MCP servers, I implement the session persistence mechanism, which maintains session states, allowing agents to seamlessly switch tool invocations across different task executions without re-establishing connections.

\section{Algorithms}

\subsection{ReAct Based Action Strategy}

\Cref{algo:action_strategy} is designed to dynamically determine the next move of an agent based on its prior trajectory and available tool set. 

\begin{algorithm}[H]
\caption{ReAct Based Action Strategy}
\algorithmicrequire User request \(r\), Trajectory \(j\), Current state \(s\), Tools \(t\) \\
\algorithmicensure Next move (t'', a')
\begin{algorithmic}[1]
\State Extract events related to \(r\) from \(j\) and \(s\) in chronological order, denoted as \(e\)

\If{\(r\) is completed according to \(e\)}
    \State return success \Comment{Goal achieved, reaching the final state}
\Else
    \State Analyze and extract the remaining subtasks, denoted as \(u\)

    \If{\(t\) contains tools capable of completing \(u\)}
        \State Select a subset of relevant tools \(t'\)
    \Else
        \State return failure \Comment{Goal determined to be unachievable, reaching the final state}
    \EndIf

    \State Plan the next action \(m\) based on \(e\)
    \State Select the tool \(t''\) from \(t'\) for executing \(m\)
    \State Generate arguments \(a'\) required for \(t''\) according to \(m\)
    \State return \(t''\), \(a'\)
\EndIf
\end{algorithmic}
\label{algo:action_strategy}
\end{algorithm}

\noindent
This approach enables the agent to adaptively adjust its actions in response to changing task requirements and environmental feedback, ensuring efficient progress toward the ultimate goal. By leveraging chain-of-thought (CoT) reasoning and tool matching analysis, the algorithm enhances the agent's decision-making capabilities and ensures robust task execution. The algorithm consists of several key steps, including relevant event extraction, task completion status determination, tool matching analysis, next move estimation, and action execution with state updates. Each step is carefully designed to optimize the agent's performance in complex and dynamic environments. The main steps of the algorithm are as follows:

\paragraph{Relevant Event Extraction}
Extract events related to the task goal from the trajectory. This step uses a language model to identify and extract relevant events, which are then sorted chronologically.

\paragraph{Task Completion Status Determination}
Analyze the extracted events to determine whether the task goal has been achieved. This step uses chain-of-thought (CoT) to optimize the accuracy of the judgment.

\paragraph{Tool Matching Analysis}
Analyze the agent's tool set to determine if there are suitable tools to advance the task goal. This step also uses CoT to ensure the judgment is strictly based on the extracted events.

\paragraph{Next Move Estimation and Tool Selection}
Estimate the next move based on the prior trajectory and task goal, select the appropriate tool, and generate the corresponding arguments.

\paragraph{Action Execution and State Update}
Execute the next move just estimated, obtain the result and environmental feedback, then update the agent's trajectory.

\subsection{Timely Abandonment Strategy}

\Cref{algo:abandonment_strategy} is designed to dynamically control the agent's task execution steps, allowing it to abandon tasks when they become unproductive or excessively resource-consuming.

\begin{algorithm}[H]
\caption{Timely Abandonment Strategy}
\algorithmicrequire \\
\begin{tabular}{l}
User request \( r \); \\
Abandonment probability \( p \); \\
Penalty coefficient \( \beta \); \\
Step estimator \( E \); \\
Next move estimator \( N \); \\
Executor \( X \);
\end{tabular} \\
\algorithmicensure Task result
\begin{algorithmic}[1]
\State \( s \gets E(r) \) \Comment{Estimate required steps using the step estimator}
\State \( c \gets 0 \) \Comment{Initialize step count}
\While{True}
    \State \( m \gets N(r, ...) \) \Comment{Estimate the next action}
    \State \( X(m) \) \Comment{Execute the next action}
    \If{\( r \) is completed}
        \State \textbf{return} success \Comment{Task successfully completed}
    \ElsIf{\( r \) is confirmed to be unattainable}
        \State \textbf{return} failure \Comment{Task failed}
    \EndIf
    \If{\( c > s \)} \Comment{Current steps exceed estimated steps}
        \State $\text{rand} \sim \text{Uniform}(0,1)$
        \If{\( \text{rand} > p \)} \Comment{Continue the task with probability \( 1 - p \)}
            \State \( p \gets (\beta \times p) \mod 1 \) \Comment{Apply penalty}
        \Else \Comment{Abandon the task with probability \( p \)}
            \State \textbf{return} failure \Comment{Abandoned, task failed}
        \EndIf
    \EndIf
    \State \( c ++ \) \Comment{Increment step count}
\EndWhile
\end{algorithmic}
\label{algo:abandonment_strategy}
\end{algorithm}

\noindent
This algorithm introduces two hyperparameters: \textbf{abandonment probability} $p$ and \textbf{penalty coefficient} $\beta$, which together regulate the agent's tendency to abandon tasks. 

\paragraph{Abandonment Probability $p$}

$p$ represents the probability that the agent will abandon the task when the current steps exceed the estimated steps. A higher $p$ makes the agent more likely to abandon tasks early, reflecting a more cautious behavior. A lower $p$ encourages the agent to persist longer, reflecting a more exploratory behavior.

\paragraph{Penalty Coefficient $\beta$}

$\beta$ is used to amplify the abandonment probability $p$ when the agent continues executing the task beyond  estimated steps. A higher $\beta$ increases the penalty more rapidly, forcing the agent to abandon tasks sooner. A lower $\beta$ allows the agent to continue exploring for longer before abandoning the task.

\paragraph{How $p$ and $\beta$ Work Together}

When an agent's current steps exceed the estimated steps, it enters a probabilistic decision-making phase. If the agent continues (with probability 1 - \textit{p}), the abandonment probability $p$ is penalized with \lcnamecref{eq:penalty} \ref{eq:penalty}, which increases the likelihood of abandonment in subsequent steps.

\begin{equation}
p = (\beta \times p) \mod 1
\label{eq:penalty}
\end{equation}

The algorithm consists of several key steps, including step estimation, action execution, task completion status determination, and probabilistic abandonment. This ensures that the agent dynamically balances exploration and caution, adapting its behavior based on task complexity and environmental feedback. 

\subsection{Memory Storage and Sharing Mechanism}

In multi-agent collaboration scenarios, memory storage and sharing mechanisms are crucial for achieving information synchronization and effective collaboration among agents. The following sections detail the multi-agent memory sharing mechanism implemented in this paper from three aspects: memory storage, memory update, and memory transfer.

\paragraph{Memory Storage}

In this mechanism, each agent is designed with its own memory storage structure to save the experience and knowledge accumulated during task execution. Memories are stored in the form of an ordered dictionary, a data structure that preserves the insertion order of memories, facilitating subsequent retrieval and processing in chronological order. The content of each memory includes timestamps, agent identifiers, action executions, and action summaries. Timestamps indicate when the memory was created, agent identifiers specify which agent generated the memory, action summaries concisely describe the core content of the action and its feedback, and action executions detail the actions taken by the agent and their associated parameters. This structured storage approach ensures that memories are well-organized and readable, providing a foundation for subsequent memory transfer and fusion.

\paragraph{Memory Update}

After each action, the agent records the result as a new memory and adds it to its memory storage. When estimating the next move, the agent retrieves its action trajectory and current state from memory to guide its subsequent actions. Once a task is completed or a termination condition is met, the agent resets its memory to prepare for the next task.

\paragraph{Memory Transfer}

Memory transfer is a critical step in multi-agent collaboration. When an agent needs assistance from other agents or needs to delegate a task, it passes its memories to the target agent. This allows the target agent to understand the current task status and existing experience of the source agent. Upon receiving the memories, the target agent merges them with its existing memories. This mechanism enables rapid information sharing among agents, avoiding redundant exploration and enhancing system efficiency. For example, if Agent A encounters a complex task and lacks specific knowledge or capabilities, it can transfer its task context and accumulated experience to Agent B via the memory transfer mechanism. With this information, Agent B can better understand the task requirements and effectively assist Agent A in completing the task.

\section{Experiment}

The primary objective of this experiment is to evaluate the performance of three frameworks: autono (the proposed framework), autogen, and langchain, in handling tasks of varying complexity. Specifically, the experiment aims to assess their adaptability and robustness as task complexity increases. Based on task complexity, the experimental tasks are categorized into three types:

\begin{enumerate}
    \item \textbf{Single-Step Tasks:} Simple tasks that can be completed with a single action.
    \item \textbf{Multi-Step Tasks:} Tasks that require a sequence of actions to complete.
    \item \textbf{Multi-Step Tasks with Possible Failures:} The most complex category, these tasks involve actions that may fail during execution, necessitating error correction and retry mechanisms. This category simulates real-world production environments with uncertainties and failures, posing higher requirements for algorithm robustness and adaptability.
\end{enumerate}

To comprehensively and objectively evaluate the performance of each framework, success rate is selected as the evaluation metric. Success rate is calculated as the percentage of tasks executed successfully as requested relative to the total number of tasks. The success rate of each framework is computed for each task type to facilitate an intuitive comparison of their performance differences. To ensure the reproducibility of the experiment, the experimental environment has been pushed to a public GitHub repository at \url{https://github.com/vortezwohl/experiment-03-22-2025}. Anyone can use this repository to reproduce the experiment.

\subsection{Experimental Results and Analysis}

\begin{table}[h]
\centering
\begin{adjustbox}{width=0.9\textwidth}
\begin{tabular}{|l|l|l|l|l|l|}
\hline
\textbf{Framework} & \textbf{Version} & \textbf{Model} & \textbf{Single-Step Tasks} & \textbf{Multi-Step Tasks} & \textbf{Multi-Step Tasks with Possible Failures} \\ \hline
autono & 1.0.0 & GPT-4o-mini & 96.7\% & 100\% & 76.7\% \\ \hline
autono & 1.0.0 & Qwen-plus & 100\% & 96.7\% & 93.3\% \\ \hline
autono & 1.0.0 & DeepSeek-v3 & 100\% & 100\% & 93.3\% \\ \hline
autogen & 0.4.9.2 & GPT-4o-mini & 90\% & 53.3\% & 3.3\% \\ \hline
autogen & 0.4.9.2 & Qwen-plus & 90\% & 0\% & 3.3\% \\ \hline
autogen & 0.4.9.2 & DeepSeek-v3 & N/A & N/A & N/A \\ \hline
langchain & 0.3.21 & GPT-4o-mini & 73.3\% & 13.3\% & 10\% \\ \hline
langchain & 0.3.21 & Qwen-plus & 73.3\% & 13.3\% & 13.3\% \\ \hline
langchain & 0.3.21 & DeepSeek-v3 & 76.7\% & 13.3\% & 6.7\% \\ \hline
\end{tabular}
\end{adjustbox}
\caption{Experimental Results}
\end{table}

\paragraph{Single-Step Tasks}

The results indicate that autono performs exceptionally well in single-step tasks. With GPT-4o-mini as the thought engine, it achieves a success rate of 96.7\%. When using Qwen-plus and DeepSeek-v3, the success rate reaches 100\%. This demonstrates that autono is highly stable and effective in handling simple tasks. In contrast, autogen shows slightly inferior performance in single-step tasks, achieving a 90\% success rate with both GPT-4o-mini and Qwen-plus. DeepSeek-v3 is unsupported in autogen, making evaluation impossible (marked as N/A). Langchain achieves success rates of 73.3\% (GPT-4o-mini), 73.3\% (Qwen-plus), and 76.7\% (DeepSeek-v3) in single-step tasks. While langchain can complete most single-step tasks, its success rate is significantly lower than that of autono and autogen.

\paragraph{Multi-Step Tasks}

In multi-step tasks, autono continues to lead, achieving success rates of 100\% (GPT-4o-mini), 96.7\% (Qwen-plus), and 100\% (DeepSeek-v3). This highlights its robust capability in handling complex tasks. Autogen's performance in multi-step tasks is notably inconsistent. With GPT-4o-mini, the success rate is 53.3\%, while with Qwen-plus, it drops to 0\%. DeepSeek-v3 is unsupported. This indicates that autogen's compatibility with third-party models is suboptimal, and it struggles to effectively execute complex tasks. Langchain's success rate in multi-step tasks is uniformly low at 13.3\% across all three models, reflecting significant limitations in handling complex task execution scenarios.

\paragraph{Multi-Step Tasks with Possible Failures}

For multi-step tasks with possible failures, autono remains highly effective. With GPT-4o-mini, it achieves a 76.7\% success rate, and 93.3\% with both Qwen-plus and DeepSeek-v3. This underscores its strong adaptability and robustness, as it can automatically correct errors and leverage past experiences to ensure task success. Autogen continues to underperform in these tasks, achieving only a 3.3\% success rate with both GPT-4o-mini and Qwen-plus. DeepSeek-v3 is unsupported. This further highlights autogen's limitations in handling complex and uncertain tasks, lacking effective error-handling mechanisms and retry logic. Langchain achieves success rates of 10\% (GPT-4o-mini), 13.3\% (Qwen-plus), and 6.7\% (DeepSeek-v3) in multi-step tasks with failures, indicating insufficient robustness in handling complex and uncertain tasks.

\subsection{Experimental Conclusion}

Overall, autono demonstrates superior performance across all task types, particularly in multi-step and multi-step tasks with possible failures. This demonstrates its superior adaptability and robustness in handling complex tasks, underscoring its advantages in complex task processing and validating the effectiveness of its ReAct-based action strategy and timely abandonment strategy, and meeting diverse scenario requirements and providing reliable solutions for developers. Autogen performs adequately in single-step tasks but sees a significant drop in success rate in multi-step and multi-step tasks with possible failures, especially when using non-OpenAI third-party models. Langchain consistently shows lower success rates across all task types, particularly struggling with multi-step and multi-step with possible failures, revealing notable limitations in multi-step execution, adaptability, and robustness. Different models exhibit high performance within autono, indicating excellent compatibility with mainstream models.

\section{Conclusion and Future Work}

The paper proposes a highly robust autonomous agent framework based on the ReAct paradigm, designed to solve complex tasks through ReAct-based action strategies and timely abandonment strategies. The framework's key innovations include: ReAct-based action strategies, timely abandonment strategies, and compatibility with the MCP protocol. Experimental results demonstrate that the framework achieves high success rates in single-step tasks, multi-step tasks, and multi-step tasks with potential failures, showcasing strong adaptability and robustness in complex and uncertain environments.

The limitations of the framework are primarily reflected in the following aspects: First, the robustness of the framework relies heavily on ReAct-based action strategies and timely abandonment strategies, which may still be insufficient to handle extreme complexity or dynamically changing environments. Second, while the multi-agent collaboration mechanism supports memory sharing, it does not deeply explore how to optimize communication efficiency and coordination among agents, which could become a bottleneck in large-scale multi-agent systems. Additionally, the experimental section focuses mainly on task success rates, but does not sufficiently address the optimization of task execution efficiency and resource consumption.

Future work could explore several directions: First, further optimizing the timely abandonment strategy to enable smarter judgments of task feasibility and reduce unnecessary resource waste. Second, investigating more efficient communication and coordination mechanisms for multi-agent systems to support large-scale complex task collaboration. Third, expanding the framework's applicability, particularly in real-time task scenarios, to validate its performance. Fourth, integrating reinforcement learning techniques to enhance the agents' adaptability and decision-making efficiency. Overall, the framework proposed in this paper provides a new perspective for designing intelligent agents in complex tasks, but there is still room for improvement, and future research can build on this foundation to further refine the framework.

\bibliographystyle{unsrt}
\bibliography{ref}

\begin{thebibliography}{10}

\bibitem{vaswani2017attention}
Ashish Vaswani, Noam Shazeer, Niki Parmar, Jakob Uszkoreit, Llion Jones, Aidan~N. Gomez, Lukasz Kaiser, and Illia Polosukhin.
\newblock Attention is all you need.
\newblock {\em arXiv preprint}, 2017.

\bibitem{devlin2018bert}
Jacob Devlin, Ming-Wei Chang, Kenton Lee, and Kristina Toutanova.
\newblock Bert: Pre-training of deep bidirectional transformers for language understanding.
\newblock {\em arXiv preprint}, 2018.

\bibitem{wei2022cot}
Jason Wei, Xuezhi Wang, Dale Schuurmans, Maarten Bosma, Brian Ichter, Fei Xia, Ed~H. Chi, Quoc~V. Le, and Denny Zhou.
\newblock Chain-of-thought prompting elicits reasoning in large language models.
\newblock In {\em Conference on Neural Information Processing Systems (NeurIPS)}, 2022.

\bibitem{yao2022react}
Shunyu Yao, Jeffrey Zhao, Dian Yu, Nan Du, Izhak Shafran, Karthik Narasimhan, and Yuan Cao.
\newblock React: Synergizing reasoning and acting in language models.
\newblock {\em arXiv preprint}, 2022.

\bibitem{talebirad2023multi}
Yashar Talebirad and Amirhossein Nadiri.
\newblock Multi-agent collaboration: Harnessing the power of intelligent llm agents.
\newblock {\em arXiv preprint}, 2023.

\bibitem{huang2024llmplan}
Xu~Huang, Weiwen Liu, Xiaolong Chen, Xingmei Wang, Hao Wang, Defu Lian, Yasheng Wang, Ruiming Tang, and Enhong Chen.
\newblock Understanding the planning of llm agents: A survey.
\newblock {\em arXiv preprint}, 2024.

\bibitem{huang2023hallucination}
Lei~Huang et~al.
\newblock A survey on hallucination in large language models: Principles, taxonomy, challenges, and open questions.
\newblock {\em arXiv preprint}, 2023.

\bibitem{hou2025mcp}
Xinyi Hou, Yanjie Zhao, Shenao Wang, and Haoyu Wang.
\newblock Model context protocol (mcp): Landscape, security threats, and future research directions.
\newblock {\em arXiv preprint}, 2025.

\bibitem{10147172}
Morteza Bahrami, Muharram Mansoorizadeh, and Hassan Khotanlou.
\newblock Few-shot learning with prompting methods.
\newblock In {\em 2023 6th International Conference on Pattern Recognition and Image Analysis (IPRIA)}, pages 1--5, 2023.

\bibitem{wu2023autogen}
Qingyun Wu, Gagan Bansal, Jieyu Zhang, Yiran Wu, Beibin Li, Erkang Zhu, Li~Jiang, Xiaoyun Zhang, Shaokun Zhang, Jiale Liu, et~al.
\newblock Autogen: Enabling next-gen llm applications via multi-agent conversation.
\newblock {\em arXiv preprint}, 2023.

\bibitem{CrewAI_crewAI_2024}
{CrewAI Team}.
\newblock {CrewAI: Framework for orchestrating role-playing, autonomous AI agents}.
\newblock \url{https://github.com/crewAIInc/crewAI}, March 2023.

\bibitem{chase2022langChain}
Harrison Chase.
\newblock {LangChain: Build context-aware reasoning applications}.
\newblock \url{https://github.com/langchain-ai/langchain}, October 2022.

\bibitem{OpenAI_Swarm_2023}
OpenAI~Solution Team.
\newblock {Swarm: Educational framework exploring ergonomic, lightweight multi-agent orchestration}.
\newblock \url{https://github.com/openai/swarm}, October 2024.

\bibitem{fourney2024magentic}
Adam Fourney, Gagan Bansal, Hussein Mozannar, Cheng Tan, Eduardo Salinas, Erkang Zhu, Friederike Niedtner, Grace Proebsting, Griffin Bassman, Jack Gerrits, et~al.
\newblock Magentic-one: A generalist multi-agent system for solving complex tasks.
\newblock {\em arXiv preprint}, 2024.

\end{thebibliography}
\end{document}